\documentclass{emulateapj}
\usepackage{amsmath}
\input{epsf}
\usepackage{graphicx}
\usepackage{epstopdf}
\submitted{Accepted for publication in ApJ on October 29, 2016}

\newcommand{\at}[2][]{#1|_{#2}}

\shorttitle{Spiral arms in galaxies orbiting a cluster}

\shortauthors{Semczuk et al.}

\begin{document}

\title{Tidal origin of spiral arms in galaxies orbiting a cluster}

\author{Marcin Semczuk\altaffilmark{1,2}, Ewa L. {\L}okas\altaffilmark{1} and Andr\'es del Pino\altaffilmark{1}}

\altaffiltext{1}{Nicolaus Copernicus Astronomical Center, Polish Academy of Sciences, Bartycka 18, 00-716 Warsaw, Poland}
\altaffiltext{2}{Warsaw University Observatory, Al. Ujazdowskie 4, 00-478 Warsaw, Poland}

\begin{abstract}
One of the scenarios for the formation of grand-design spiral arms in disky galaxies involves their interactions with a
satellite or another galaxy. Here we consider another possibility, where the perturbation is instead due to the
potential of a galaxy cluster. Using $N$-body simulations we investigate the formation and evolution of spiral arms
in a Milky Way-like galaxy orbiting a Virgo-like cluster. The galaxy is placed on a few orbits of different
size but similar eccentricity and its evolution is followed for 10 Gyr. The tidally induced, two-armed, approximately
logarithmic spiral structure forms on each of them during the pericenter passages. The spiral arms dissipate and wind
up with time, to be triggered again at the next pericenter passage. We confirm this transient and recurrent nature of
the arms by analyzing the time evolution of the pitch angle and the arm strength. We find that the strongest arms are
formed on the tightest orbit, however they wind up rather quickly and are disturbed by another pericenter passage. The
arms on the most extended orbit, which we analyze in more detail, wind up slowly and survive for the longest time.
Measurements of the pattern speed of the arms indicate that they are kinematic density waves. We attempt a comparison
with observations by selecting grand-design spiral galaxies in the Virgo cluster. Among those, we find nine examples
bearing no signs of recent interactions or the presence of companions. For three of them we present close structural
analogues among our simulated spiral galaxies.

\end{abstract}

\keywords{
galaxies: clusters: general --- galaxies: evolution --- galaxies: interactions
--- galaxies: kinematics and dynamics --- galaxies: spiral --- galaxies: structure }

\section{Introduction}

According to the Galaxy Zoo project, spiral galaxies contribute around two thirds of all galaxies in the Local Universe
(Lintott et al. 2011; Willett et al. 2013). While the spiral structure is very common and appealing, the mechanism
underlying its origin is still not well understood. Several theories aim to explain the nature of spiral arms in
disky galaxies, however none of them is believed to be complete and universally applicable (for a review see Dobbs \&
Baba 2014).

In their seminal work Lin \& Shu (1964) proposed that the spiral arms are non-material, quasi-stationary density waves
that rotate with the fixed pattern speed. This theory was later developed reaching a mature state (Bertin et al. 1989;
Bertin \& Lin 1996), however one of its main predictions, namely the long lifetime of the spiral pattern, was
difficult to reproduce in numerical simulations and little observational evidence for it was available (Sellwood 2011).
Numerical studies typically found that spiral arms appear to be transient and short-lived. This kind of spiral arms
are often referred to as `dynamic spirals' and seem to be triggered by the swing amplified perturbations or noise
in the stellar disk (e.g. Sellwood \& Carlberg 1984; Fujii et al. 2011; Grand et al. 2012; Baba et al. 2013; D'Onghia
et al. 2013). While these arms are dynamic and wind up fast, the recurrent mechanism of the perturbations can maintain
the spiral structure in the galaxies for cosmological timescales (Fujii et al. 2011). Dynamic spirals in the
simulations tend to have flocculent, multi-arm morphologies. Only recently Saha \& Elmegreen (2016) succeeded in
creating long-lived ($\sim$ 5 Gyr) spiral wave modes. To accomplish that, they performed simulations of a galaxy
with high values of the Toomre $Q$ parameter in the inner region provided by the bulge, which was interpreted as a
barrier reflecting the wave and ensuring its long survival.

In a recent paper, Hart et al. (2016) showed that the fraction of galaxies with the arm number $m=2$ is greater in
regions of higher density, which indicates that they are of a different origin than those with $m>2$. It
is well known that some of the grand-design, two-armed spirals originate from tidal interactions with other
galaxies. This scenario is in fact the one most established observationally given the evidence of
galaxies like M51 interacting with its nearby companion NGC 5195. A list of $\sim 20$ two-armed galaxies interacting
with a companion of a different size was recently published by Gunthardt et al. (2016).

From the theoretical point of view, the tidally induced spiral structure in interacting galaxies was first seen in
the seminal work by Holmberg (1941). Later on, following the development of numerical calculations more detailed
studies of the interacting galaxies were performed (Toomre \& Toomre 1972; Eneev et al. 1973). Nowadays, most
efforts focus on simulations of a normal-size galaxy interacting with a smaller companion. These studies include pure
$N$-body approaches (Oh et al. 2008, 2015) as well as hydrodynamical simulations (Dobbs 2011; Struck et al. 2011;
Pettitt et al. 2016). There have also been attempts aiming to reproduce particular observed systems like M51 (Salo \&
Laurikainen 2000; Dobbs et al. 2010) or M81 (Yun 1999). From these works the following picture emerges: a flying-by
companion is inducing a tidal bridge-tail structure in the main galaxy, that later winds up to transform into
grand-design spiral arms. These arms keep winding up and dissipating which manifests itself in the decrease of the
pitch angle and the strength of the arms in time. Such a process usually takes about 1 Gyr. The pattern speed of the
arms tends to decrease with radius and follows or slightly exceeds the inner Lindblad resonance, which means that the
arms are non-material kinematic density waves.

The magnitude of the tidal perturbation can be quantified by the dimensionless parameter $S$ defined by Elmegreen et al.
(1991):
\begin{equation}            \label{defs}
	S=\bigg(\frac{M_{\mathrm{ptb}}}{M_{\mathrm{gal}}}\bigg)\bigg(
	\frac{R_{\mathrm{gal}}}{d}\bigg)^3\bigg(\frac{\Delta T}{T}\bigg),
\end{equation}
where $M_{\mathrm{ptb}}$ is the mass of the perturber, $M_{\mathrm{gal}}$, $R_{\mathrm{gal}}$ are the mass and the
characteristic size of the perturbed galaxy and $d$ is the distance between both bodies at closest approach. $\Delta
T$ is the interaction time defined as the time that the perturber needs to move over an angle of one radian around the
progenitor and $T$ is the time for stars in the outer part of the disk of the progenitor to move one radian in their
orbits, which can also be expressed as $T=(R_{\mathrm{gal}}^3/G M_{\mathrm{gal}})^{1/2}$. Elmegreen et al.
(1991), Oh et al. (2013, 2015) and Pettitt et al. (2016) explored different values of $S$ and found that spiral arms
can be triggered by a smaller companion with the parameter $S$ in the range $0.01<S<0.25$. Equation (\ref{defs}) shows
that a very similar tidal perturbation can be obtained from a companion dwarf on a tight orbit and a bigger body on an
appropriately wider orbit. In this work we use $N$-body simulations to investigate a scenario in which the perturber
is a Virgo-like cluster and spiral arms are induced in a Milky Way-like galaxy orbiting around it.
In this rescaled configuration including a much larger perturber the range of values of $S$ we find
(assuming $M_{\mathrm{ptb}}$ is the mass of the cluster enclosed within $d$) are very similar,
$0.09<S<0.14$.

Interactions between the cluster potential and the orbiting galaxies were previously discussed by Merritt (1984), where
it was shown that the spherical galaxy may be tidally truncated by the cluster. Later $N$-body (Byrd \& Valtonen 1990)
and restricted three-body (Valluri 1993) calculations demonstrated that the cluster's tidal field can induce transient
two-armed spiral structure in disky galaxies (as well as a bar in the case of Byrd \& Valtonen 1990). Recently Bialas et
al. (2015) considered tidal interactions between the cluster and infalling galaxies, however the main focus of that
work was to investigate the process of galaxy harassment (Moore et al. 1996, 1998) and its ability to transform disky
galaxies into other morphological types. In this paper we focus on the formation and evolution of
spiral arms that are tidally excited in a galaxy interacting with the cluster using $N$-body simulations.

The paper is organized as follows. In Section 2 we present the simulations used in this study. In Section 3 we discuss
in detail the properties of the spiral arms in the galaxy on the most extended orbit, focusing on their shape,
amplitude and velocity. Section 4 compares spiral arms induced on different orbits using some of the quantitative
methods described in Section 3. In Section 5 we attempt to place the scenario in the observational context. Finally,
Section 6 provides the discussion and summary of the most important findings of the paper.

\section{The simulations}

We use $N$-body simulations of the Milky Way-like galaxy orbiting a Virgo-like cluster to investigate the formation and
evolution of tidally induced spiral arms. In this work we use the same simulations as were described in \L{}okas
et al. (2016) and used to study tidally induced bars. Initial conditions for the simulations were generated with the
procedures described in Widrow \& Dubinski (2005) and Widrow et al. (2008). The Virgo cluster was approximated as a
Navarro-Frenk-White (NFW; Navarro et al. 1997) dark matter halo of $10^6$ particles with parameters estimated by
McLaughlin (1999) and Comerford \& Natarajan (2007), namely the virial mass $M_{\mathrm{C}}=5.4 \times
10^{14}\;\mathrm{M}_{\odot}$ and the concentration $c=3.8$.

\begin{figure}
\begin{center}
\includegraphics[width=225pt]{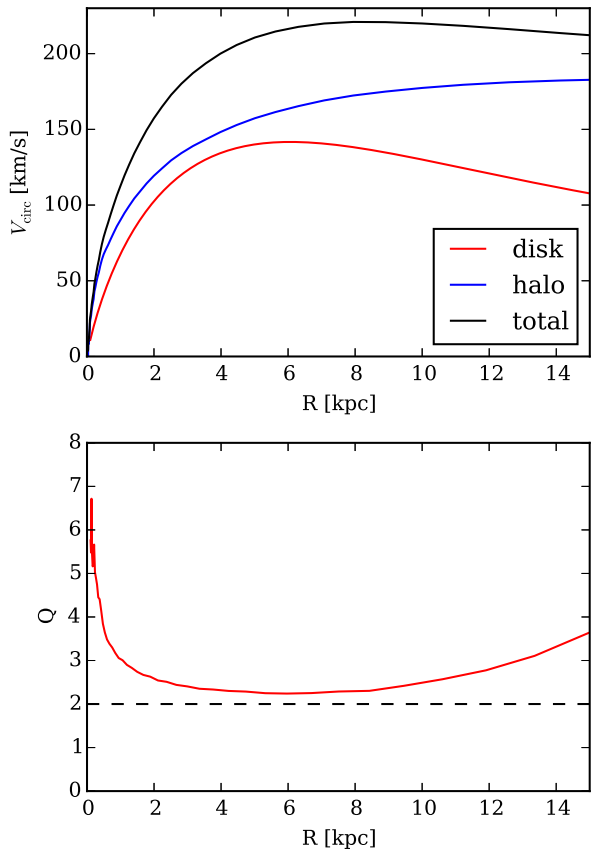}
\end{center}
\caption{ Upper panel: the initial rotation curve of the progenitor galaxy.
Lower panel: the initial radial profile of the Toomre stability parameter $Q$.}
\label{rot}
\end{figure}

The progenitor galaxy was modelled as a two-component system similar to the Milky Way. The two components were an NFW
dark matter halo and an exponential stellar disk, each made of $10^6$ particles. The model was similar to the model MWb
of Widrow \& Dubinski (2005). The dark matter halo had a virial mass $M_{\mathrm{H}}=7.7\times
10^{11}\;\mathrm{M}_{\odot}$ and concentration $c=27$, while the disk had a mass
$M_{\mathrm{D}}=3.4\times10^{10}\;\mathrm{M}_{\odot}$, the scale-length $R_{\mathrm{D}}=2.82$ kpc and thickness
$z_{\mathrm{D}}=0.44$ kpc. Both components were smoothly cut off at large radii. The upper panel of
Figure~\ref{rot} shows the initial rotation curve of the galaxy and the contributions from the two components. The
initial conditions were such that the Toomre parameter was $Q>2.1$ at all radii (see the lower panel of
Figure~\ref{rot}), preventing the formation of strong morphological structures when evolving the galaxy in isolation
for a few Gyr.

The progenitor galaxy was placed on four typical, eccentric orbits in the Virgo cluster with a typical apo- to
pericenter distance ratio $D_{\mathrm{apo}}/D_{\mathrm{peri}}=5$ (Ghigna et al. 1998). All the orbits were coplanar and
prograde with respect to the galaxy disk, with apo- and pericentric distances summarized in Table 1. The simulations we
will refer to here as O1-O4 correspond to S1-S4, respectively, in \L{}okas et al. (2016). The orbital periods for
simulations O1-O4 were 1.3, 1.9, 2.5 and 3.7 Gyr.

\begin{table}
\begin{center}
\caption{Orbital parameters of the simulations}
\begin{tabular}{lccl}
\hline
\hline
Simulation  & $D_{\mathrm{apo}}$ [Mpc] & $D_{\mathrm{peri}}$ [Mpc] & Line color \\
\hline
O1 & \    0.5  & \    0.1  & \ \ red \\
O2 & \ \, 0.75 & \ \, 0.15 & \ \ green \\
O3 & \    1.0  & \    0.2  & \ \ cyan \\
O4 & \    1.5  & \    0.3  & \ \ blue \\
\hline
\label{initial}
\end{tabular}
\end{center}
\end{table}

The evolution was followed for 10 Gyr with the GADGET-2 $N$-body code (Springel et al. 2001; Springel 2005) with
outputs saved every 0.05 Gyr. The adopted softening scale for the halo of the Virgo cluster was
$\epsilon_{\mathrm{C}}=14$ kpc while for the halo and disk of the progenitor $\epsilon_{\mathrm{H}}=0.7$ kpc and
$\epsilon_{\mathrm{D}}=0.1$ kpc.

\section{Formation and evolution of the spiral structure}

The spiral arms form on each orbit in the simulations, however we find that the most persistent arms occur for the most
extended orbit O4. Therefore, in this section we will focus on the case O4 and discuss the comparison between the
arms forming on different orbits in Section 4. We choose this case because the longevity of the arms and
the relatively weak bar (see \L{}okas et al. 2016) allow for more precise quantitative studies, but the general
behavior described in this section applies to all orbits after some rescaling. First results
concerning orbit O4 were already presented in Semczuk \& {\L}okas (2015).

\begin{figure*}
\begin{center}
\includegraphics[width=500pt]{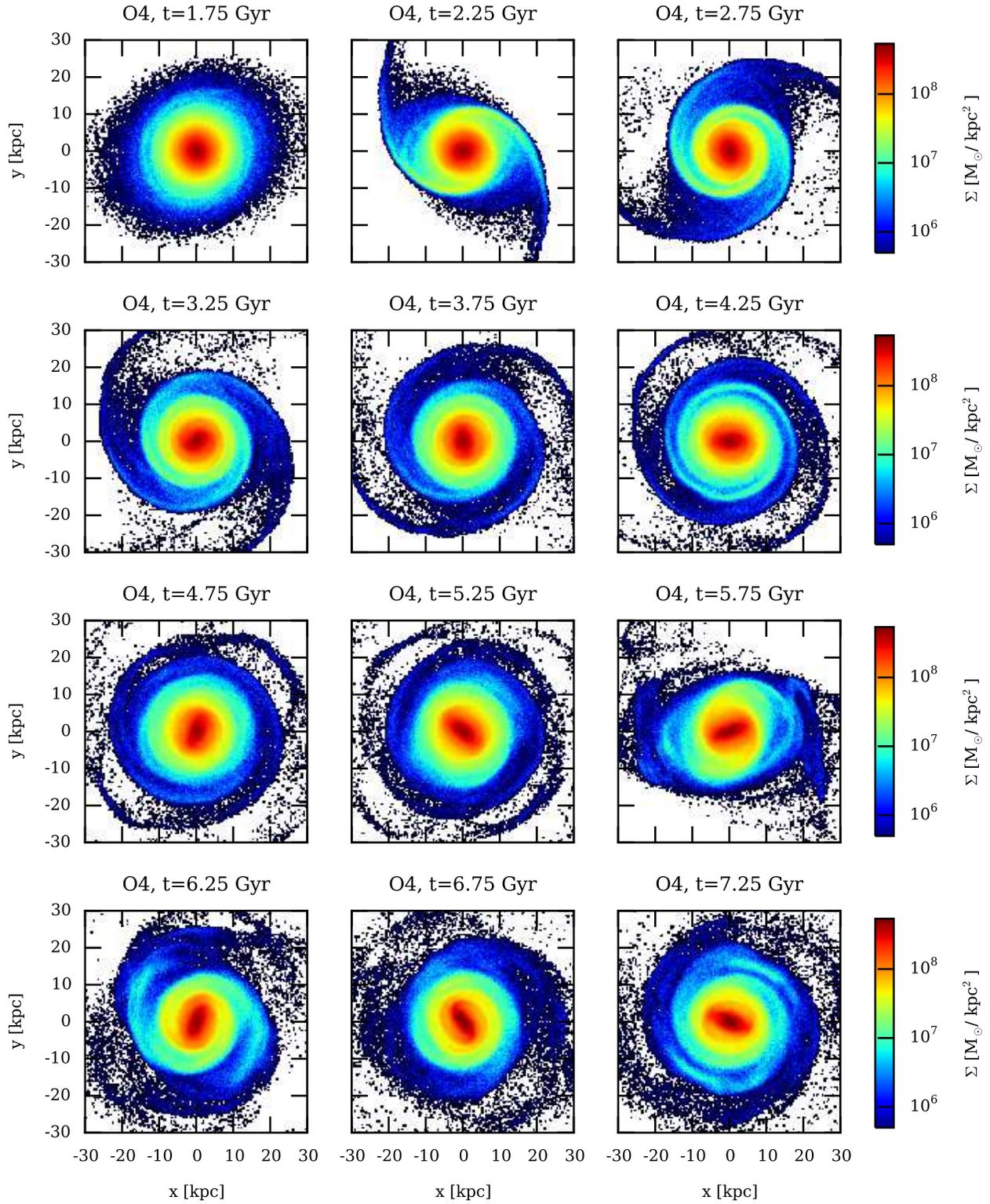}
\caption{Face-on views of the surface density distribution of stars $\Sigma$ in the disk for orbit O4 at different
times. The first and second pericenter passages occurred at 1.9 Gyr and 5.4 Gyr, respectively.}
\label{snaps}
\end{center}
\end{figure*}

\begin{figure*}
\begin{center}
\includegraphics[width=500pt]{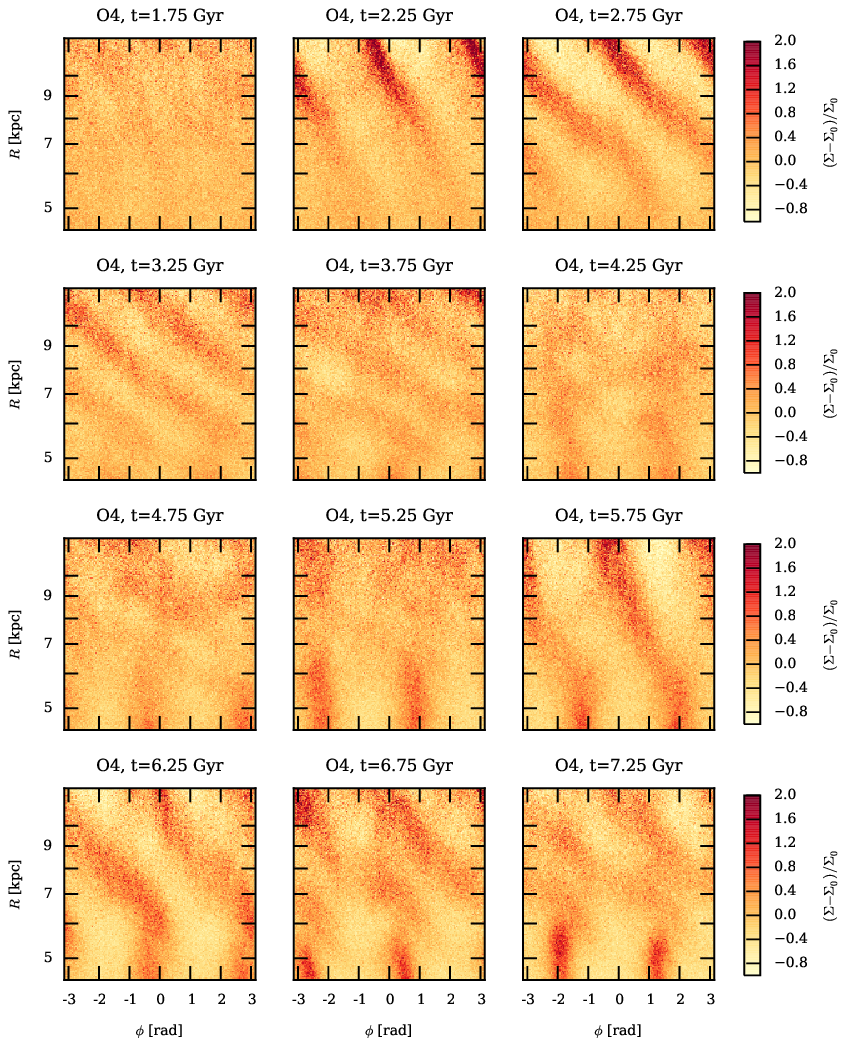}
\caption{Face-on views of the perturbed density of stars $(\Sigma-\Sigma_0)/\Sigma_0$ in the $\phi$ - $\ln R$ plane for
orbit O4 at different times, the same as in Figure~\ref{snaps}. The first and second pericenter passages occurred at 1.9
Gyr and 5.4 Gyr, respectively.}
\label{phiR}
\end{center}
\end{figure*}

\subsection{Overview}

We find that the formation of spiral arms is triggered by the pericenter passages. During the pericenters tidal
forces of the cluster cause the stars from the galaxy disk to form tidal tails. However, most of the stars in
the tails are still bound to the progenitor, hence the structure winds up towards the center of the galaxy to form
spiral arms. Later on the arms keep winding up and dissipate to be triggered again during the next pericenter
passage. This dynamic and recurrent behavior is well seen in the plots comprising Figure~\ref{snaps}. From the face-on
views of the surface density of the stars $\Sigma$ in Figure~\ref{snaps} we can also infer that the induced structure
is as expected two-armed and of grand-design type.

The shape of the grand-design spiral arms can be often approximated with the logarithmic spiral. In the plots of
Figure~\ref{phiR} we show the time evolution of the perturbed density defined as $(\Sigma-\Sigma_0)/\Sigma_0$ (where
$\Sigma_0$ is the initial face-on density distribution of the stars) in the $\phi$ - $\ln R$ plane, where $(\phi,\;R)$
are polar coordinates in the plane of the disk (see also e.g. Oh et al. 2008, 2015). If the spiral arms were perfect
logarithmic spirals, the overdensities corresponding to the arms would have the shape of
straight lines in the plots of Figure~\ref{phiR}. As we can see, these lines are not perfectly straight, however as a
first approximation we can treat these arms as logarithmic. Figure~\ref{phiR} also confirms the winding up of the arms
and their transient and recurrent nature. The lower plots of Figure~\ref{phiR} (for $t>$ 5 Gyr) also reveal the
formation of the bar in the form of vertical overdense regions at smaller radii.

\subsection{Fourier analysis}

\begin{figure}
\begin{center}
\includegraphics[width=200pt]{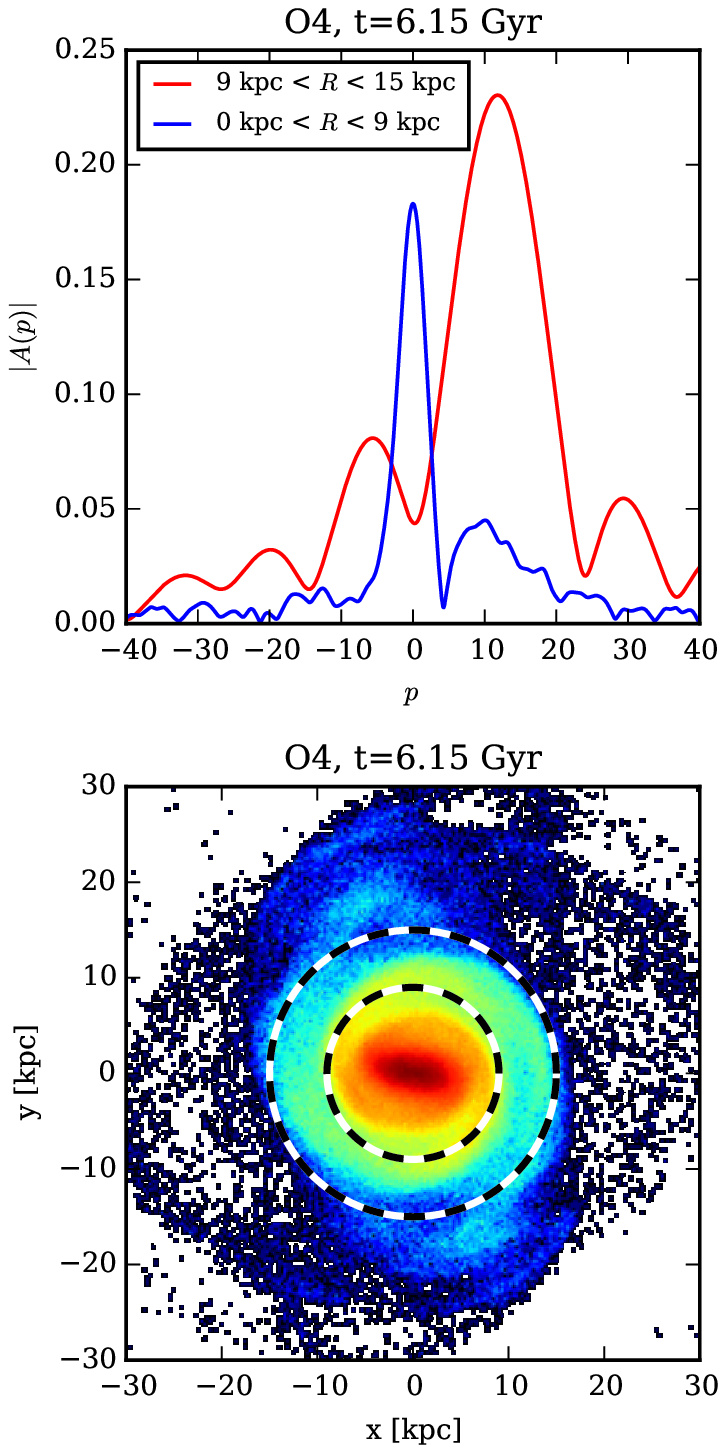}
\end{center}
\caption{Upper panel: example of $|A(p)|$ calculated in two regions, the inner and outer one. Lower panel: surface
density of stars with dashed circles marking two regions for which $|A(p)|$ was calculated in the upper panel.}
\label{Ap}
\end{figure}

As demonstrated by Figure~\ref{phiR}, the tidally induced spiral arms in our simulations can be approximated as
logarithmic spirals. We use this fact to expand the surface distribution of stars in logarithmic spirals as
discussed in e.g. Sellwood \& Athanassoula (1986) and Oh et al. (2008, 2015). The expansion is given by the
formula
\begin{equation}
	A(m,p)=\frac{1}{N_s}\Sigma_j \exp [i(m \phi_j+p \ln R_j)],
\end{equation}
where $N_s$ is the number of stars, $(\phi_j,\;R_j)$ are the polar coordinates of the $j$-th star, $m$ is the number of
spiral arms (here we will only consider $m=2$) and $p$ is a parameter related to the pitch angle
$\alpha$. We calculated the function $|A(2,p)|\equiv|A(p)|$ in the fixed ring 9 kpc $\leq$ $R$ $\leq$ 15 kpc, and then
found the $p_{\mathrm{max}}$ that maximizes this function to obtain the pitch angle using the relation $\tan \alpha=
2/p_{\mathrm{max}}$. We chose this range of radii for the ring to make sure that the bar will not influence the
results. We justify this choice in Figure~\ref{Ap}, which demonstrates that $|A(p)|$ calculated in $R$ $\leq$
9 kpc has a maximum around $p \simeq 0$. This means that the bar can be interpreted as spiral arms with the pitch
angle $\alpha \simeq 90^{\circ}$ and therefore contaminate our measurements. In Section 4, where we compare the results
for different orbits we will pick the ring even further from the center because the bar is there longer.

\begin{figure}
\begin{center}
\includegraphics[width=200pt]{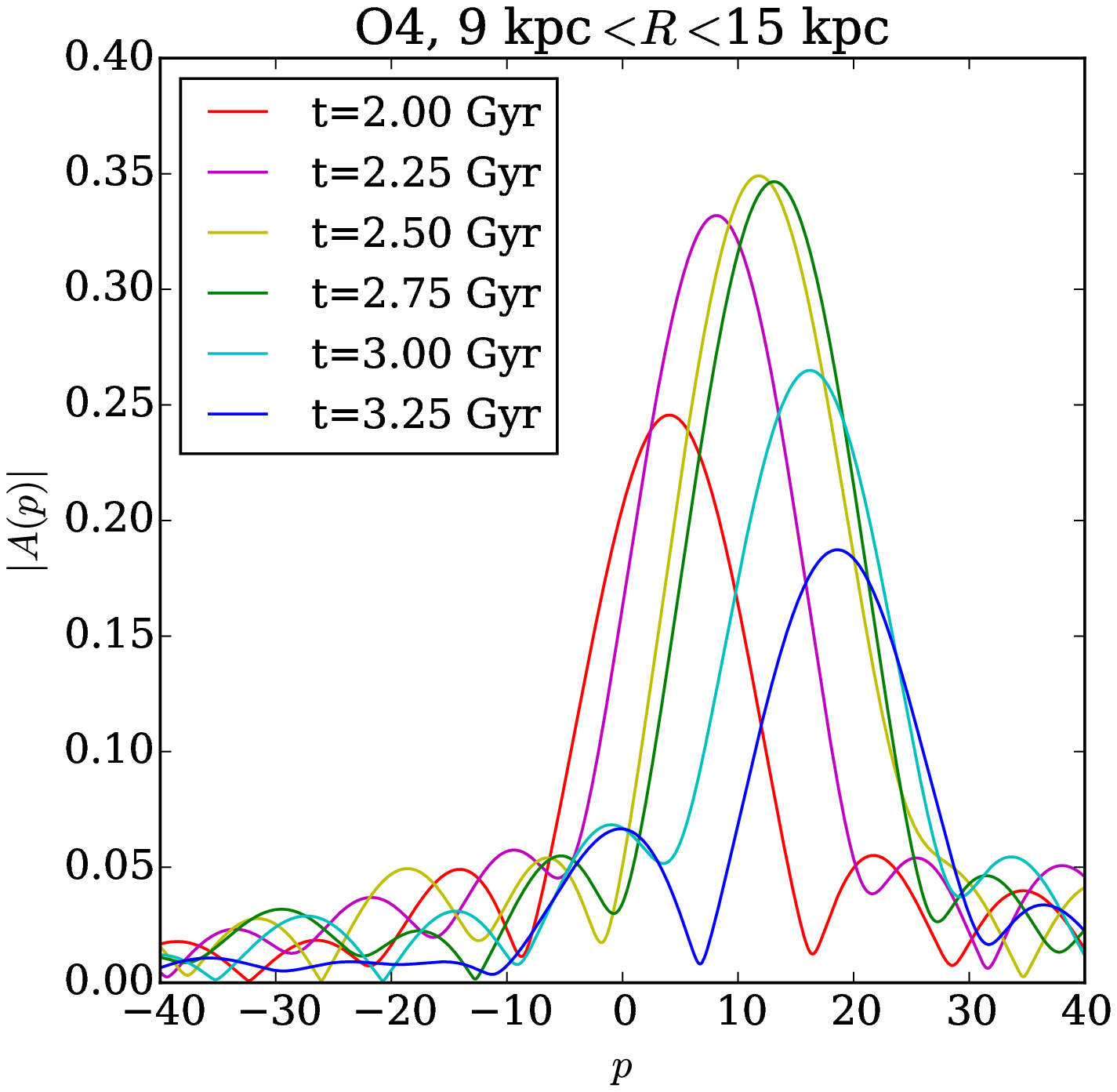}
\end{center}
\caption{Time evolution of $|A(p)|$ calculated in the fixed ring. Note that the pericenter passage occurred at 1.9 Gyr.}
\label{ap_evo}
\end{figure}

\begin{figure}
\begin{center}
\includegraphics[width=240pt]{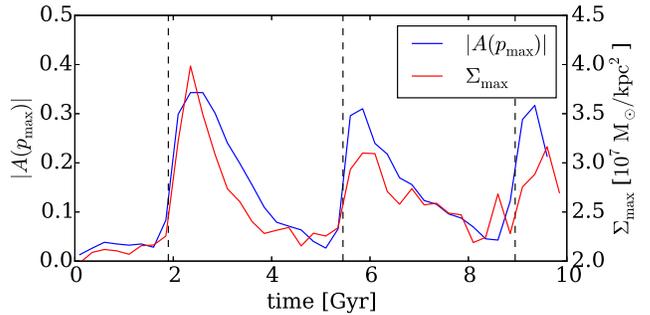}
\end{center}
\caption{Time dependence of the arm strength $|A(p_{\mathrm{max}})|$ (blue line) calculated in the ring 9 kpc $\leq$
$R$ $\leq$ 15 kpc and the maximum arm surface density $\Sigma_{\mathrm{max}}$ (red line) calculated in the annuli
of 10.2$\pm$0.3 kpc. Both measurements were made for orbit O4. Dashed vertical lines indicate pericenter passages.}
\label{str}
\end{figure}

\begin{figure}
\begin{center}
\includegraphics[width=225pt]{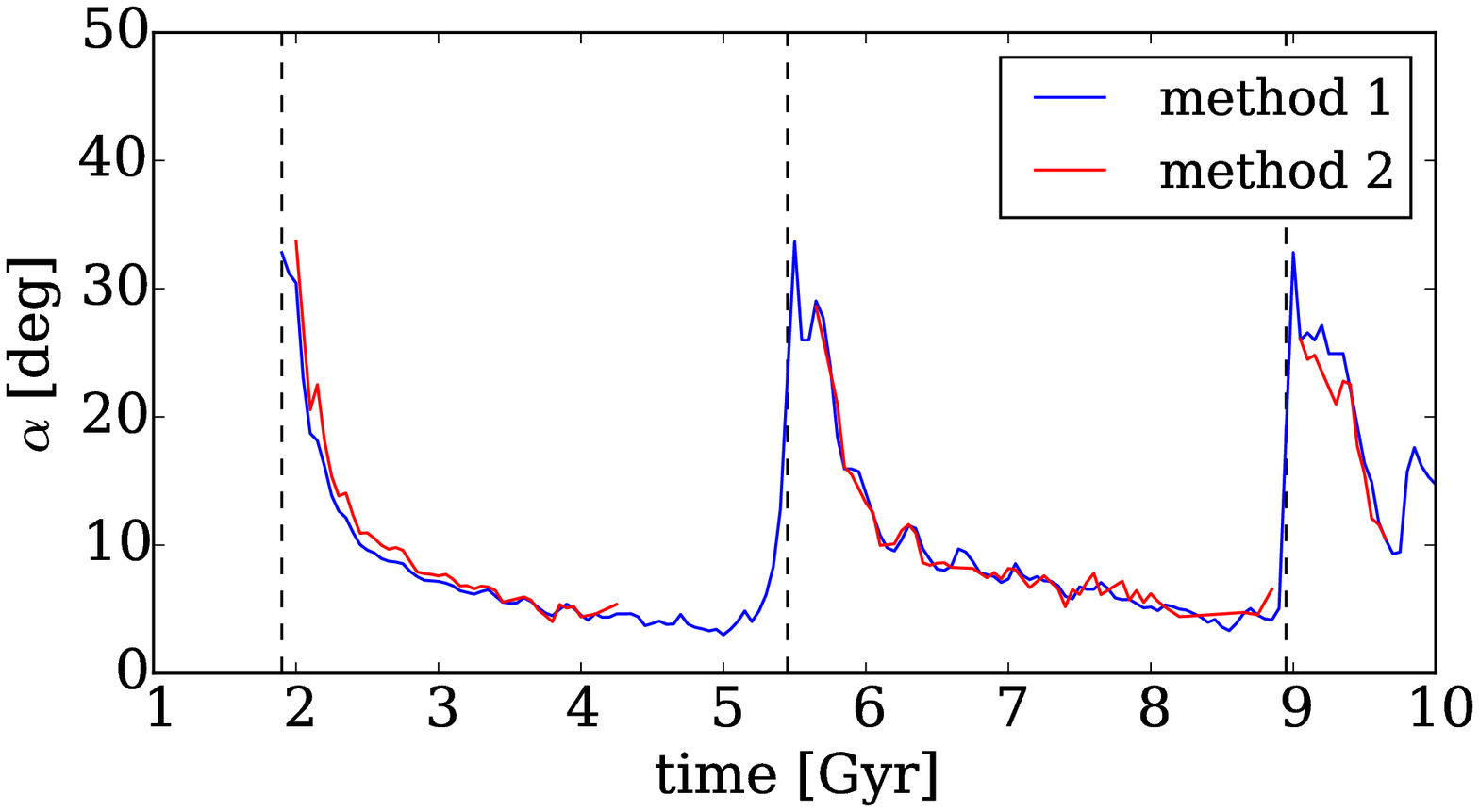}
\end{center}
\caption{Time dependence of the pitch angle $\alpha$ for orbit O4, calculated in the ring 9 kpc $\leq$ $R$ $\leq$
15 kpc with two different methods. Dashed vertical lines indicate pericenter passages.}
\label{pitch}
\end{figure}

The time evolution of $|A(p)|$ calculated in the chosen ring shortly after the first pericenter passage on orbit O4 is
shown in Figure~\ref{ap_evo}. The value of $p_{\mathrm{max}}$ is changing very rapidly toward higher values which
corresponds to the winding up of the arms and the decrease of the pitch angle $\alpha$. Note that the value of
$|A(p_{\mathrm{max}})|$ is also changing non-randomly with time. We define $|A(p_{\mathrm{max}})|$ as the parameter
measuring the arm strength and plot its time evolution in Figure~\ref{str}. The time dependence of the pitch
angle $\alpha$ is shown in Figure~\ref{pitch} labeled as method 1.

Right after the pericenter passage the pitch angle has a value $\alpha \simeq 30^{\circ}$ and afterwards it
exponentially decreases to values below $10^{\circ}$. The same behavior occurs after the next pericenter. The
decrease of the pitch angle confirms that the spiral arms wind up between the pericenters as was already seen in
Figures~\ref{snaps} and \ref{phiR}. Note that the pitch angles in the range of $10^{\circ}$ to
$30^{\circ}$ are very realistic values that are indeed measured in observed galaxies (Binney \& Tremaine 1987; Ma 2002).

The arm strength $|A(p_{\mathrm{max}})|$ behaves similarly to the pitch angle: it has highest values around the
pericenter and then exponentially decreases until the next pericenter. This confirms that the arms are strongest right
after they are formed and with time they dissolve. However, there is clearly a difference between the evolution of
$|A(p_{\mathrm{max}})|$ and $\alpha$: the peaks of the arm strength are shifted to $\sim0.5$ Gyr after the
pericenters. This is due to the fact that the tidal features formed during the passage are strongest in the outer parts
of the disk and they need time to wind up and migrate into the ring in which we do the measurements.

\subsection{The pitch angle from surface density fits}

In the previous subsection we applied one method to derive the pitch angle of the spiral arms. Here we use another
approach  (which is also based on the assumption that the arms can be described as logarithmic spirals) to confirm our
findings. The method is very similar to the one presented in Grand et al. (2013) and it consists of fitting logarithmic
spirals to the surface density distribution of the stars $\Sigma$.

\begin{figure}
\begin{center}
\includegraphics[width=220pt]{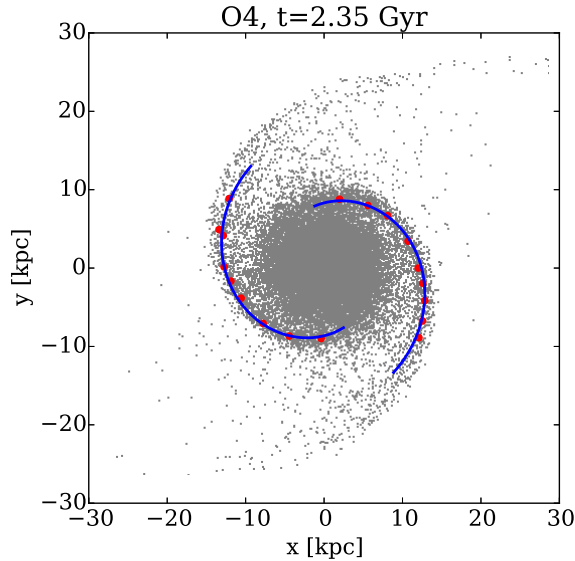}
\end{center}
\caption{Face-on view of a random subsample of disk particles (gray). Red points are the selected maxima
$\phi_{\mathrm{max},\;j}$ of the surface density at given radii $R_j$. Blue lines are the logarithmic spirals fitted
to the red points (see subsection 3.3).}
\label{scat}
\end{figure}

\begin{figure}
\begin{center}
\includegraphics[width=220pt]{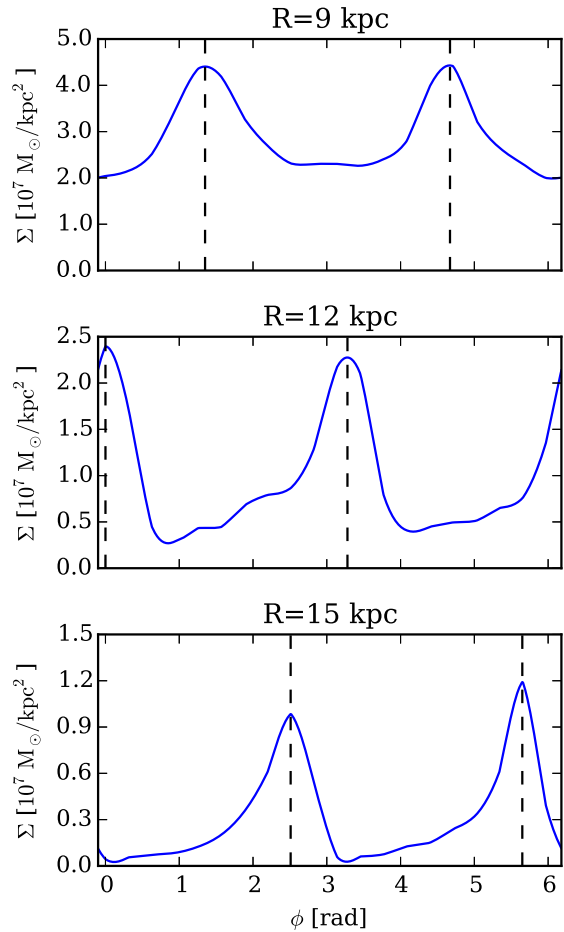}
\end{center}
\caption{The dependence of the surface density of stars $\Sigma$ on the azimuthal angle $\phi$ at three different
radii: 9 kpc, 12 kpc and 15 kpc. Dashed vertical lines indicate the maxima found numerically. Results are
plotted for orbit O4 at $t=2.35$ Gyr. The maxima correspond to the red points at the same radii in Figure~\ref{scat}.}
\label{maxima}
\end{figure}

First, we find $\Sigma$ in the polar coordinates ($\phi$, $R$) and then we look at a given radius $R_j$ for local maxima
$\phi_{\mathrm{max},\;j}$ corresponding to the two arms. We select $R_j$ from the same range 9 kpc $\leq$ $R$ $\leq$ 15
kpc as in the previous subsection so that the results of the two methods are comparable. Next, using the least squares
method, we fit the logarithmic spiral
\begin{equation}
	\phi=B \ln R + C
\end{equation}
to the two sets of points ($\phi_{\mathrm{max},\;j}, R_j$). The procedure is illustrated in Figure~\ref{scat} where we
plot the points used for the fit, the fitted logarithmic spirals and a subsample of stars of the simulated galaxy.
A few examples of the plots of $\Sigma(\phi)$ at fixed radii with marked maxima are presented in Figure~\ref{maxima}.

The pitch angle $\alpha$ of each arm is given by $\tan \alpha=1/|B|$. The time dependence of the average pitch angle of
the two arms is plotted in Figure~\ref{pitch} labeled as method 2. We find that the two methods are in very good
agreement. We note that method 2 requires more parameters so when using these procedures to automatically deal with
a large number of simulation outputs, method 1 seems more straightforward to apply.

\subsection{The maximum arm surface density}

In subsection 3.2 we have introduced $|A(p_{\mathrm{max}})|$ as an indicator of the strength of the spiral arms. Here
we apply a different approach that was recently used in Few et al. (2016). This approach is very simple and consists
of tracking the value of the maximum surface density $\Sigma_{\mathrm{max}}\equiv\Sigma(\phi_{\mathrm{max}})$ at a
fixed radius as a proxy of the arm strength. The profiles of the surface density at different radii were already shown
in Figure~\ref{maxima}. We choose as a measure of the arm strength the mean of the two values of the maxima at
$10.2\pm0.3$ kpc (since it shows least noise) and present the time evolution of this quantity in Figure~\ref{str}. For
the other radii the behavior of $\Sigma_{\mathrm{max}}$ is very similar.
However, the adopted distance from the center of the galaxy affects the values of the maxima due to the growing bar
at smaller radii. In general, the evolution of $\Sigma_{\mathrm{max}}$ is very similar to the
evolution of $|A(p_{\mathrm{max}})|$: the peaks occur at $\sim0.5$ Gyr after the pericenter passage and then the value
exponentially decreases until the next pericenter. The agreement between the two methods of measuring the arm strength
confirms our findings concerning the recurrent and transient evolution of the spiral structure.

\subsection{The pattern speed}

The pattern speed of the spiral arms is an essential parameter concerning the nature of the spiral arms. According
to the quasi-stationary density wave theory introduced by Lin \& Shu (1964) the spiral pattern should rotate like a
rigid body with a fixed, constant pattern speed $\Omega_{\mathrm{p}}$. However, if the arms are kinematic density waves
their rotation should follow the inner Lindblad resonance i.e. $\Omega_{\mathrm{p}}(R)=\Omega(R)-\kappa(R)/2$. Finally,
if the spiral arms are material they should rotate in the same way as the stars in the disk,
$\Omega_{\mathrm{p}}(R)=\Omega(R)$ (see e.g. Dobbs \& Baba 2014). To find the pattern speed of the arms in our
case we use two methods out of many available in the literature, which are most often applied to simulations.

\begin{figure}
\begin{center}
\includegraphics[width=200pt]{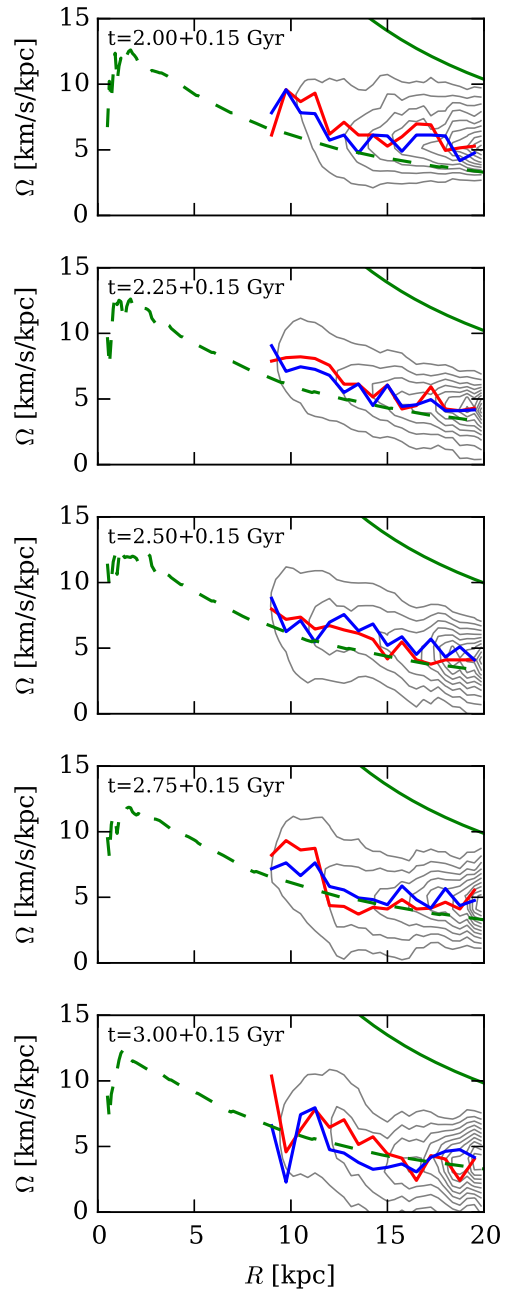}
\end{center}
\caption{Contours of the cross-correlation $C(R,\phi,t)$ for orbit O4 after the first pericenter passage at $t=$1.9 Gyr.
Contours are spaced by 10\% of the maximum value of $C(R, \phi, t)$.
Red and blue lines indicate the pattern speed of each arm measured separately. Solid green lines mark $\Omega$ of the
stars while dashed green lines correspond to the inner Lindblad resonance $\Omega - \kappa/2$.}
\label{pattern}
\end{figure}

The first method we apply was introduced by Oh et al. (2008, 2015). It consists of calculating the normalized
cross-correlation of the perturbed surface density at two different times separated by $\Delta t$
\begin{equation}
	C(R, \phi, t)=\frac{1}{\Sigma_0 (R)^2} \int ^{2 \pi} _0 \delta \Sigma (R, \xi, t) \delta \Sigma
	(R, \xi+\phi, t+\Delta t) d \xi,
\end{equation}
where $\delta \Sigma=\Sigma-\Sigma_0$ and $\xi$ is a polar angular coordinate over which the expression is integrated.
We choose $\Delta t$=0.15 Gyr (which is $3\times0.05$ Gyr, with 0.05 Gyr being the time step
between our saved simulation outputs) because the arms formed on orbit O4 seem to be relatively slow. Once $C(R, \phi,
t)$ is calculated, we obtain the pattern speed at a given radius by finding $\phi_{\mathrm{max}}$ that maximizes
the cross-correlation. The pattern speed is given by the relation $\Omega_{\mathrm{p}}(R,t)=\phi_{\mathrm{max}}/\Delta
t$. The contours of the cross-correlation indicating the locus of the maximum are plotted in Figure~\ref{pattern}.

The second method we use was discussed e.g. in Dobbs (2011) and seems more straightforward. It consists of finding the
maximum of the density (here we use the surface density $\Sigma$) of stars at a given radius in polar coordinates, at
two different epochs. The pattern speed at a fixed radius is expressed by a simple formula
\begin{equation}
	\Omega_{\mathrm{p}}(t)=\frac{\phi(\Sigma_{\mathrm{max}})\at[]{t+\Delta t}
	-\phi(\Sigma_{\mathrm{max}})\at[]{t}}{\Delta t}.
\end{equation}
Here for the same reason we also use $\Delta t$=0.15 Gyr. The pattern speeds calculated with this approach for both
arms are plotted with red and blue lines in Figure~\ref{pattern} (note that the color coding does not follow the same
arm in different plots).

The comparison of the lines and contours in the plots of Figure~\ref{pattern} confirms that the two methods give
consistent results. Both arms show similar radial dependence, which also overlaps with the contours of the
cross-correlation. Although the two tidally induced arms are expected to have slightly different pattern speeds due to
the asymmetry of the process (Dobbs 2011), we do not find any systematic offset between them, probably because of the
relatively mild tidal forces. We note however that our measurements are a bit noisy and done some time after the arms
are formed.

Clearly the arms are not quasi-stationary density waves because the pattern speed decreases with radius.
In addition, the pattern speed profile lies very close to the inner Lindblad resonance indicating that the arms are
kinematic density waves. Over the 1 Gyr for which the measurements are shown in Figure~\ref{pattern}, the range of
values of the pattern speed is approximately the same and only varies radially from 10 to 4 km
s$^{-1}$ kpc$^{-1}$. The value of $\Omega_{\mathrm{p}}\simeq 6$ km s$^{-1}$ kpc$^{-1}$ in the outer parts of the disk
soon after the first pericenter passage is close to the angular velocity of the progenitor on its orbit around the
Virgo-like cluster, $\Omega_{\mathrm{orb}}\simeq 6.2$ km s$^{-1}$ kpc$^{-1}$. This concurrence confirms the tidal
origin of the arms and was previously noted in the literature (Oh et al. 2008, 2015).

\section{Comparison between the orbits}

\subsection{General properties}
As mentioned at the beginning of the Section 3, we find that the general behavior of the tidally induced arms is
qualitatively similar for all orbits considered in this work but still some dependence on the orbit is present. To
illustrate these differences we use the approach based on the expansion into logarithmic spirals (described in section
3.2) due to its simplicity, however we also show some qualitative differences in plots similar to those in
Figures~\ref{snaps} and \ref{phiR} for the different orbits at the approximately similar evolutionary stages.

\begin{figure}
\begin{center}
\includegraphics[width=230pt]{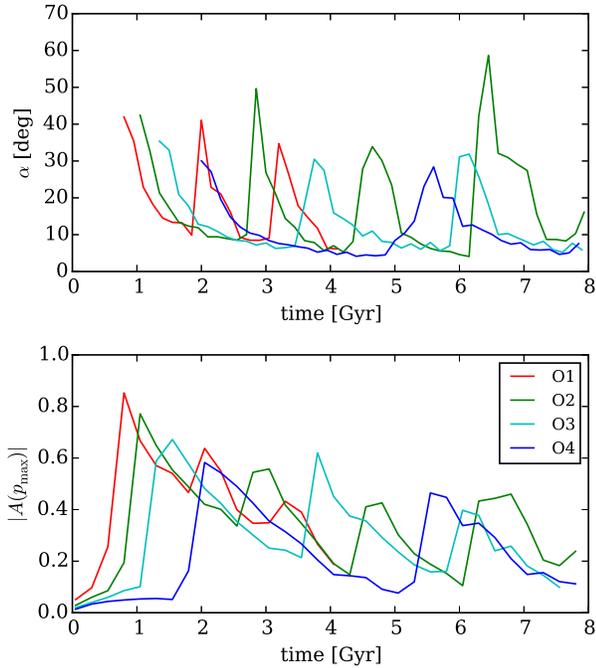}
\end{center}
\caption{Upper panel: the time evolution of the pitch angle $\alpha$ for all orbits measured in the ring 12 kpc $\leq$
$R$ $\leq$ 17 kpc. Lower panel: the time evolution of the arm strength $|A(p_{\mathrm{max}})|$ for all orbits measured
in the same region.}
\label{comp}
\end{figure}

{\L}okas et al. (2016) demonstrated that the strongest and most extended bar forms for the tightest orbit O1.
To make sure that this strong bar does not influence the measurements concerning the spiral arms (see Figure~\ref{Ap})
and to maintain consistency we choose to compare the results for all orbits using the same ring of 12 kpc $\leq$ $R$
$\leq$ 17 kpc. At later times in the simulations the strong bar influences $|A(p)|$ even in this far region, however
going even further away would not provide enough stellar particles to obtain smooth $|A(p)|$ functions. To avoid the
effects of the growing bar we show the time dependence of the pitch angle $\alpha$ and the arm strength
$|A(p_{\mathrm{max}})|$ in Figure~\ref{comp} for orbits O2-O4 only for the first 8 Gyr and for orbit O1 for the first 4
Gyr.

\begin{figure*}
\begin{center}
\includegraphics[width=500pt]{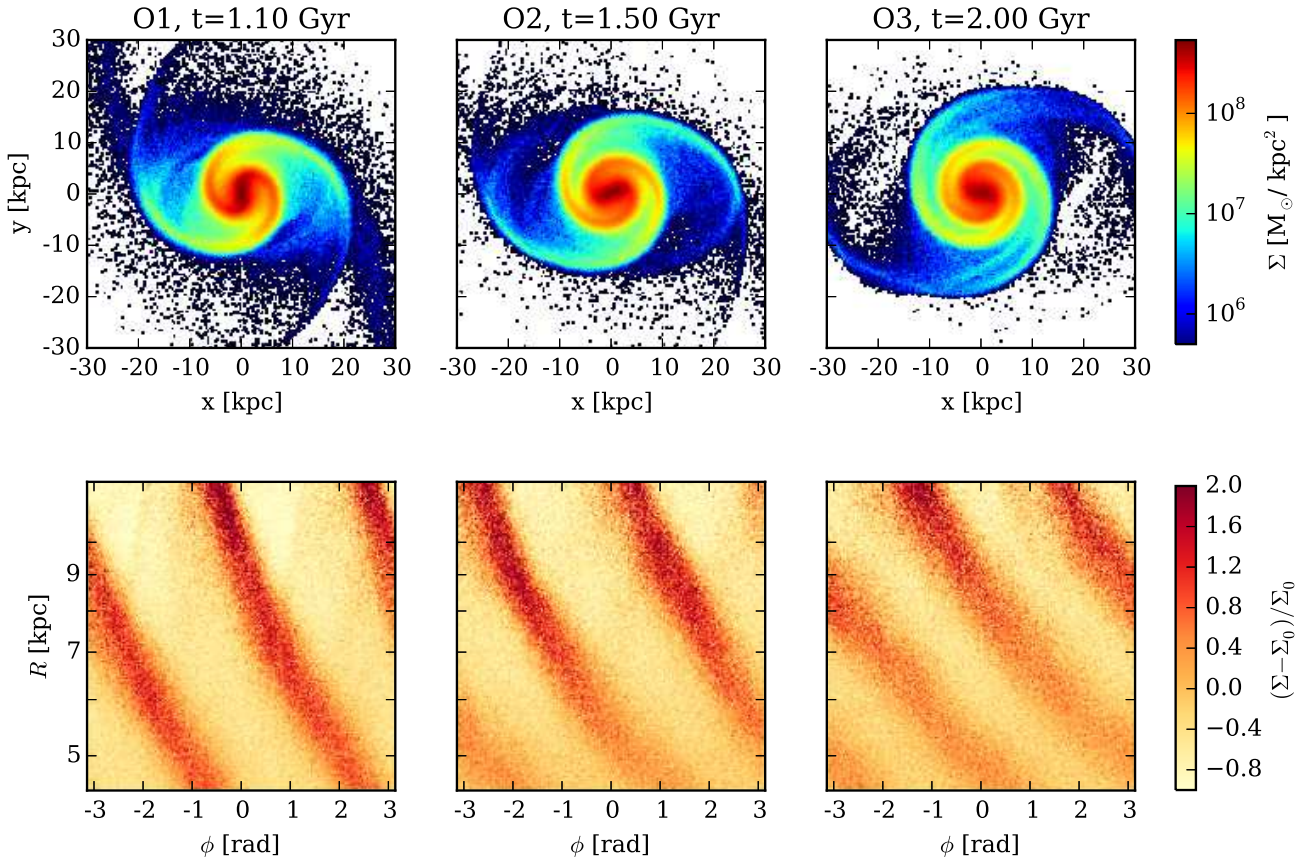}
\caption{Upper panels: face-on views of the surface density distribution of stars $\Sigma$ in the disk for orbits
O1-O3 (columns) at $t= 0.3 T_{\mathrm{orb}}$ after the first pericenter passages. Lower panels: the perturbed density
of stars $(\Sigma-\Sigma_0)/\Sigma_0$ in the $\phi$ - $\ln R$ plane for the same orbits at the same epochs.}
\label{comp2}
\end{center}
\end{figure*}

For all orbits right after the pericenter the pitch angle $\alpha$ (see the upper panel of Figure~\ref{comp}) has a
value around $30^{\circ}$-$40^{\circ}$. After that it exponentially decreases to $\sim 5^{\circ}$ and repeats the
cycle after the next pericenter passage. The timescale of this process depends on the orbit and the orbital period.
However, also the slope of the decrease depends on the orbit to some extent. It is well visible especially when
comparing orbits O1-O3 with O4: for the most extended orbit the slope is less steep. This means that the most persistent spiral arms or the ones that are winding up most slowly occur for the most extended orbit O4. However the effect of the steepness of the slope is very weak and the durability of the O4 arms is mostly
due to the long orbital period and relatively mild tidal forces.

The time dependence of the arm strength $|A(p_{\mathrm{max}})|$ (see the lower panel of Figure~\ref{comp}) also
confirms the recurrent and transient evolution of the spiral arms for all orbits. The differences between the slopes
are not well visible, however there is a clear difference in the values of $|A(p_{\mathrm{max}})|$. The tighter the
orbit, the stronger the arms are in terms of $|A(p_{\mathrm{max}})|$: it ranges from $\sim 0.85$ for O1 to $\sim
0.6$ for O4. This finding is consistent with the same dependence for the bar (\L{}okas et al. 2016). We may therefore
conclude that tidally induced (or enhanced) morphological features are stronger for tighter orbits in terms of the
Fourier coefficients.

Figure~\ref{comp2} shows face-on stellar surface density distributions (upper panels) and the perturbed density
distributions in the $\phi$ - $\ln R$ plane (lower panels) for orbits O1-O3 at $t=0.3 T_{\mathrm{orb}}$ (where
$T_{\mathrm{orb}}$ is the orbital period) after their first pericenter passages. The spiral arms and the disk in
general show some differences between the orbits at this approximately the same evolutionary stage. First, we can see
that for orbits O1 and O2 the spiral arms are peeling off from the tidal tails. It is more transparent for O1, while
for O2 this effect is visible closer to the tips of the arms. Some time after the pericenter passage the tidal-spiral
structure starts to separate: the particles that are still bound to the galaxy wind up and form the spiral arms,
while less bound particles detach from the galaxy and form the tidal tails. The same phenomenon occurs for orbit O3,
however its timescale is different and the quadrupole structure is seen later.
More information about
the spiral arms can be inferred from the lower panels of Figure~\ref{comp2}. The plots demonstrate that for the tighter
orbits the perturbed density takes higher values in the inner parts. It means that for O1 the strong arm structure
reaches deepest into the disk at this particular evolutionary stage. On the other hand, the spirals for the most
extended orbit seem slightly wider and more wound up.

\begin{figure}
\begin{center}
\includegraphics[width=200pt]{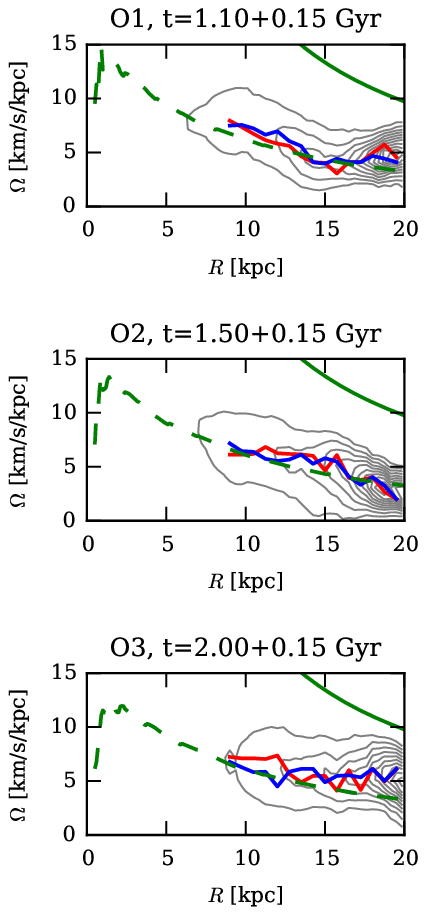}
\end{center}
\caption{Contours of the cross-correlation $C(R,\phi,t)$ for orbits O1-O3 at $t= 0.3 T_{\mathrm{orb}}$ after
the first pericenter passages. Contours are spaced by 10\% of the maximum value of $C(R, \phi, t)$. Red and blue lines
indicate the pattern speed of each arm measured separately. Solid green lines mark $\Omega$ of the stars while dashed
green lines correspond to the inner Lindblad resonance $\Omega - \kappa/2$.}
\label{pattern_comp}
\end{figure}

We also compared the pattern speed of spiral arms forming on orbits O1-O3 at the same epochs for which the density maps
in Figure~\ref{comp2} were made. The results, obtained with the two methods described in section 3.5, are
presented in Figure~\ref{pattern_comp}. We find that the ranges of the pattern speed for all orbits are very
similar. All of them also seem to follow tightly the inner Lindblad resonance. For orbits O1-O2 the radial dependence
seems to be a bit steeper than for O3. However, after analyzing the time evolution of $\Omega_{\mathrm{p}}(R)$ we find
that this flatness in O3 is not very significant. The slope of the radial decrease of the pattern speed seems to be
rather noisy and its changes are not correlated with any particular events during the orbital evolution. For O3 this
decrease happens to be steeper before and after the epoch for which we presented the results here. We note that the
plots of Figure~\ref{pattern_comp} also confirm the agreement between the two methods of deriving the pattern speed.

\subsection{Radial displacement of the stars}
Radial migration of stars in galaxies was first discussed by Sellwood \& Binney (2002) and can be described as a process
of changing the orbital angular momentum of the stars without changing the eccentricity of their orbits. Several
authors investigated the influence of the non-axisymmetric structures like spiral arms on the radial migration of the
stars (e.g. Sellwood \& Binney 2002; Vera-Ciro et al. 2014; Martinez-Medina et al. 2016). To verify how much the orbits
of the stars are changed in our simulations we apply one of the methods discussed by Martinez-Medina et al. (2016).

\begin{figure}
\begin{center}
\includegraphics[width=220pt]{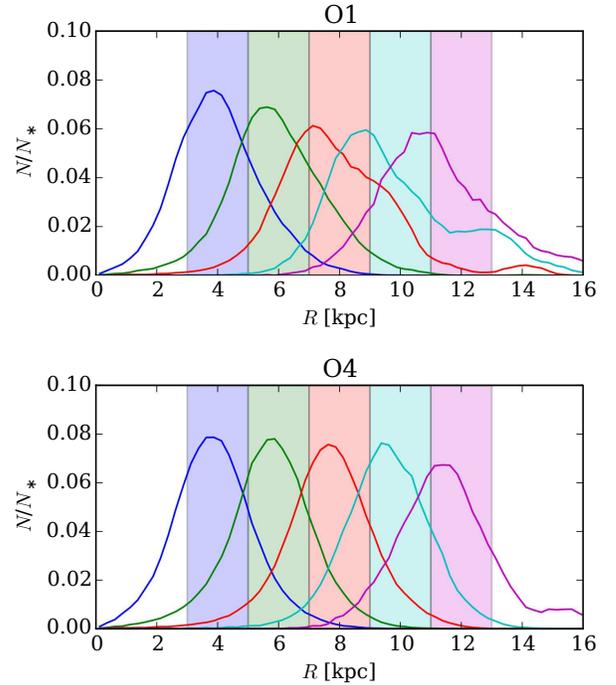}
\end{center}
\caption{Initial distribution of stellar radii (during the first apocenter, colored lines) that during the next
apocenter are located within the corresponding radial bin (shaded regions of similar color) for
two extreme orbits O1 and O4. $N$ is the initial number of stars at the given radius and $N_*$ is the total number of
stars located within each radial bin at the next apocenter.}
\label{rad_mig}
\end{figure}

In Figure~\ref{rad_mig} we show the initial (at the first apocenter) distributions of the radii of the stars (colored
lines) that later, at the second apocenter, were found in the radial bin of similar color (shaded regions). We present
these plots for the two extreme orbits O1 and O4 in order to clearly see the influence of the tidal force (for O1
the tidal parameter $S=0.14$, for O4 $S=0.09$) and the bar that is formed during the first pericenter for O1 but not for
O4. The displacement of the peak of the distribution with respect to the center of the colored bin contains the
information on whether a significant fraction of the stars migrated outwards or inwards. For both orbits in
Figure~\ref{rad_mig} we find that the further away from the center of the galaxy the stars initially were, the further
outwards are they shifted from their initial positions. This most probably excludes the bar as the driver of the radial
migration since for O1 after the first pericenter the bar size was below 5 kpc.

For orbit O1 where a greater tidal force was acting on the disk ($S=0.14$), the displacement is larger and the
distributions are not symmetric Gaussians. For the milder encounter ($S=0.09$) on orbit O4 the displacements are
smaller and the distributions may be approximated by the normal distribution. The difference between O1 and O4
suggests that the tidal force is responsible for pulling the stars outwards, since the effect is greater for the
greater force. One may still interpret the smaller shift of the radial distributions of the stars for O4  (if we assume
that the tidal force is negligible in comparison with O1) as mainly due to radial migration caused by the spiral arms as
discussed in Sellwood \& Binney (2002). However, we verified that in the case of O4 after the pericenter the orbits of
the stars become significantly more eccentric and therefore we conclude that the radial shift might be due to the mixed
effect of the mild tidal torquing and the scattering of the stars on the spiral arms (Elmegreen \& Struck 2013; 2016).

\section{Comparison with observations}

In this work we have shown that it is possible to induce the grand design spiral arms in the Milky Way-like galaxy only
by the tidal interactions with a galaxy cluster.
It remains to be investigated whether the presence of gas can significantly alter this picture.
However, it was previously shown that pure $N$-body simulations can produce spiral arms
via interactions with a satellite or a similar-sized galaxy (e.g. Toomre \& Toomre 1972; Yun 1999; Oh et al. 2008,
2015). In this study we have extended this list of the possible perturbers to include cluster-size objects and we
have measured the general properties of the induced spiral pattern. The question now is whether our setup is realistic
and whether this scenario is really taking place in the Local Universe.

In order to attempt to answer this question we searched for grand-design spiral galaxies in the Virgo cluster,
additionally imposing the condition that they do not show any signs of interaction with a satellite or another
galaxy. Our candidates were selected from three extragalactic databases:
NASA/IPAC Extragalactic Database (NED), HyperLeda (Makarov et al. 2014) and Galaxy Zoo (Lintott et al.
2011). We first performed the search for all galaxies located within the radius of 10$^{\circ}$ from the position
of M87, selecting only those with velocities in the range of $-1000$ km s$^{-1} < v < 3000$ km s$^{-1}$. This criterion
corresponds approximately to a $3\sigma$ cut in the velocity and cleans the sample of obvious interlopers. Then we
selected only those galaxies for which reliable spiral classification was available. By these we mean galaxies which
have been classified as spirals in at least two of these catalogues, giving however twice bigger weight to the
classification provided by NED. Such a selection yields a sample of 201 spiral galaxies.

\begin{table*}
\centering
\caption{Selected grand-design spiral galaxies from the Virgo cluster}
\begin{tabular}{lllcl}
\hline
\hline
Name            & \ $\alpha$ {[}deg{]} & \ $\delta$ {[}deg{]} & Morphological type\textsuperscript{a}
& Anemic\textsuperscript{b} \\ \hline
NGC 4067        & 181.0481           & 10.8544            & SA(s)b\textsuperscript{c}    & -\\
NGC 4208 (4212) & 183.914            & 13.9015            & SAc                             & no \\
NGC 4450        & 187.12346          & 17.08494           & SA(s)ab                         & yes\\
NGC 4535        & 188.58462          & \, 8.19775         & SAB(s)c                         &  no\\
M58 (NGC 4579)  & 189.43134          & 11.81819           & SAB(rs)b                        & debatable\\
M91 (NGC 4548)  & 188.86022          & 14.49634           & SB(rs)b                         & yes\\
NGC 4580        & 189.45162          & \, 5.36852         & SAB(rs)a pec                    &  no     \\
IC 3267         & 186.02303          & \, 7.04128         & SA(s)cd                         &  -\\
UGC 7133        & 182.33229          & 18.9975            & SABd                            &  -\\ \hline
\multicolumn{5}{l}{\textsuperscript{a}\footnotesize{de Vaucouleurs et al. (1991)},
\textsuperscript{b}\footnotesize{Koopmann \& Kenney (2004)}}\\
\multicolumn{5}{l}{\textsuperscript{c}\footnotesize{A bar is clearly seen in this galaxy in newer images so we would classify it as SB(s)b}}
\end{tabular}
\end{table*}

\begin{figure*}
\begin{center}
\includegraphics[width=500pt]{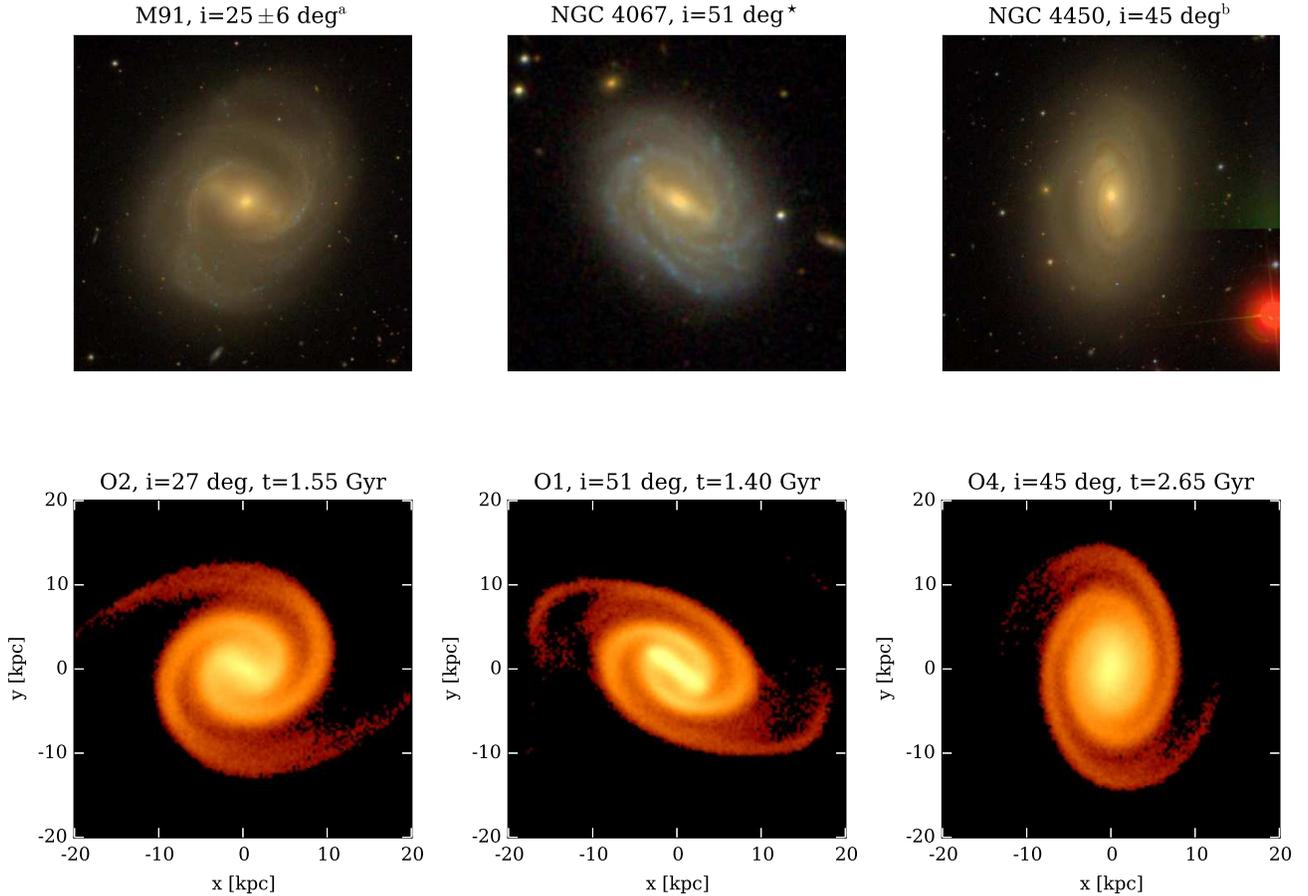}
\caption{Upper panels: SDSS images of three galaxies selected from Table 2. Lower panels: surface density maps of the
simulated galaxies modified to mimic the images of the corresponding real galaxies. The plots aim only to illustrate
the general morphological similarities between the simulated and real objects, in particular the inner parts of the
spiral arms and their connection with the bar. Inclinations of the galaxies were adopted from $^{\mathrm{a}}$Vollmer et
al. (1999) and $^{\mathrm{b}}$Cayatte et al. (1990). $^{\star}$The inclination of NGC 4067 was estimated with the
formula from Bottinelli et al. (1983) and the axis sizes from SIMBAD (Skrutskie et al. 2006).}
\label{obs}
\end{center}
\end{figure*}

Then we visually inspected the sample to look for grand-design, two-armed spiral galaxies. This `by eye' selection
yielded 24 galaxies. Afterwards we searched the literature for any signatures of past interactions with a
similar-sized galaxy or a satellite. We have excluded the objects that were classified in SIMBAD database (Wenger et
al. 2000) as a Group of Galaxies, a Pair of Galaxies or Interacting Galaxies. After this brief research we obtained a
list of 9 galaxies showing the grand-design spiral pattern and for which there is no evidence of their recent
interactions with dwarfs or other galaxies. The identifiers and the basic information about these galaxies are
summarized in Table 2. Note that our literature search was very basic and we welcome any comments concerning
the possible signatures of interactions with other galaxies for the objects listed in Table 2. One may
argue that the fact that we do not see any signs of interactions with other galaxies does not mean that there
were none in the past. It is obviously true, but the probability that all 9 galaxies were perturbed by a satellite or a
fly-by galaxy and we see no evidence of it must be low. This small sample we present is just intended to show that our
idealized scenario from the simulations is possible and the origin of the spiral arms in these galaxies may be due to
the interaction with the Virgo cluster.

While we do not find any stellar streams pointing towards a satellite or any similar evidence of interaction with
other galaxies near the objects listed in Table 2, there is some evidence from the radio observations that they
might have been interacting with the intracluster medium (ICM). In particular, M91 (also known as NGC 4548) shows
perturbations in its gaseous content (Vollmer et al. 1999) and NGC 4535 shows asymmetry in the structure of its magnetic
field (We\.zgowiec et al. 2007). These signatures point toward past ram-pressure stripping (RPS) caused by ICM,
that could only favor our scenario of interaction between the galaxies and the cluster. Still, pure ram-pressure
induced morphologies would show characteristic, asymmetric, mainly one-armed morphologies (e.g., Kenney et al. 2014),
in contrast with our tidally formed two-armed structures. The same applies to the galaxies classified as anemic
(Koopmann \& Kenney 2004): it means they have low star formation rate (SFR), that could have been decreased by the RPS.
We note however that recent simulations by Steinhauser et al. (2016) show that RPS caused by the cluster has only small
influence on the quenching.

The selected 9 galaxies do not reside in any particular region of the Virgo cluster, their spatial distribution is more
or less uniform. The majority of them have projected distances to M87 smaller than 1.5 Mpc which is the largest
apocenter in our simulations. While some of them have greater projected distances than 1.5 Mpc, they still lie within
the virial radius of the Virgo cluster ($\sim3$ Mpc), although we obviously cannot determine their
orbits. We cannot say whether galaxies on orbits with still larger apocenters would produce tidally induced spirals.
Note, however, that the Virgo cluster is not a spherically symmetric system and is known to possess a few substructures
so the galaxies that are further away from M87 may have interacted with a closer massive subcluster.

In our simulated galaxies we found that the pitch angle changes from initial $30^{\circ}$-$40^{\circ}$ down to $\sim$
5$^{\circ}$. This covers all the typical values of the pitch angle (van den Bergh 1998) corresponding to different
morphological types of the Hubble sequence. Just like Sundelius et al. (1987) we reproduced every Hubble type with
different arm winding from Sc to Sa. Morphologies of our galaxies also resemble SDSS pictures of galaxies from
the sample listed in Table 2. In Figure~\ref{obs} we compare SDSS images of three galaxies (upper panels) with
appropriately inclined and rotated surface density maps of our simulated galaxies (lower panels). The images of the
simulated galaxies in Figure~\ref{obs} have also been cut in the density below the threshold corresponding to
the surface brightness of 25 mag arcsec$^{-2}$ assuming the mass-to-light ratio of $M/L=2$ solar units. This procedure
was performed to hide the extensive tidal tail-bridge structures that may not be visible in the SDSS images due to the
limited surface brightness range.

Comparing the images in Figure~\ref{obs} we find good agreement between the shapes of the  morphological structures
in observations and simulations. Although all the maps from the simulations lack the bright core seen in the
observations, we note that our simulated galaxy did not include a bulge initially and our goal was not to reproduce
these particular galaxies. In addition, in the observed images the outer parts of the spiral arms seem to be more
tightly wound, especially for M91 and NGC 4067. Regardless of that, the shape and the thickness of the inner parts of
the spiral arms, and their connection with the bar (particularly in the case of NGC 4067) are in very good agreement.
This is even more remarkable given the fact that in the simulations we considered a general scenario and not the case
of these three galaxies.

\section{Discussion and Summary}

\subsection{Discussion}

The main findings of this paper concerning the nature of the spiral arms are in a good agreement with previous works
considering somewhat different setups, yet focusing on the tidally induced spiral structure. Oh et al. (2008) performed
2D and later 3D (Oh et al. 2015) $N$-body simulations of a disky galaxy perturbed by a companion. Recently Pettitt et
al. (2016) revisited this configuration including hydrodynamical simulations.

In these papers it has been found that the pitch angle of the spiral arms peaks briefly after the closest approach of
the companion and exponentially decreases from values $\lesssim40^{\circ}$ to $\sim 5^{\circ}$ in about 1 Gyr. We
observe a similar behavior with the same range of values and timescales in our simulations over one orbital period.
It is especially well visible for orbit O4 (Figure~\ref{pitch}) where 1 Gyr after the pericenter the pitch angle drops
to $\sim 7^{\circ}$ and then, due to the exponential nature of the curve, the decrease is very small. The timescale of the winding of the spiral arms we find here seems greater than the wind-up time of arms in a galaxy orbiting a cluster inferred from the face-on snapshots published by Byrd \& Valtonen (1990) and Valluri (1993).
The difference probably arises from the fact that in these early works the particle resolution was significantly
lower than nowadays but may also arise from different initial conditions. Pettitt et al. (2016) found some monotonic
changes in the time evolution of the pitch angle in different models. However, it is difficult to compare them with our
results concerning the different winding rate between orbit O4 and orbits O1-O3 discussed in Section 5 of this paper,
because these changes are very subtle and there are too many differences between both simulations (e.g. a different
setup, the inclusion of the gas, varying the mass of the perturber and not only the orbit).

Besides the similar time evolution of the pitch angle corresponding to the winding of the arms, also the decrease of
the arm strength found in the papers of Oh et al. (2008, 2015) and Pettitt et al. (2016) is similar to our results
when considered over one orbital period. The exponential decrease of the arm strength reflects the decay of the arms
found in each of the simulations, however it is difficult to compare the specific values due to the fact that in each
paper a different approach is applied to measure the arm strength.

Another similarity is the radial dependence of the pattern speed that follows $\Omega-\kappa/2$ or slightly exceeds
this level due to self-gravity (Oh et al. 2015). The resemblance also includes the small variability of this
radial dependence. However, the spirals discussed in this paper, induced by the cluster-like halo, seem to have lower
pattern speeds than those in the other papers, where the spiral arms are induced by the companions. We note that this
may also be a result of the specific properties of our progenitor galaxy.

Despite the very similar picture concerning the spiral arms emerging from papers by Oh et al. (2008, 2015), Pettitt et
al. (2016) and this work, one may still question the tidal origin of the arms discussed in this paper. The reason for
that may be the presence of the bar and its connection with the spiral arms. In the aforementioned papers
the bar formation was suppressed by the inclusion of the bulge component in modelling the main galaxy. In this paper,
we used the simulations that were originally designed to investigate the bar formation. However, as demonstrated by
figures 6 and 8 in {\L}okas et al. (2016), the bar for each orbit grows almost monotonically, even for the isolated
case S5, while spirals appear and decay in tight correlation with the pericenter passages and are almost non-existent
when the galaxy is evolved in isolation. Because of the fact that the oscillatory behavior in Figure~\ref{comp} does
not correlate with the bar growth, while it does with the orbital motion, we can exclude the hypothesis that the
spirals originate from the bar and confirm that they are tidally induced. We also note that while the bar is
not the driver of the spiral structure and it may appear as an obstacle in the measurements of the properties of the
spirals, the fact that it does form in our simulations provides a more complete picture of the evolution of galaxies in
clusters. Indeed, a significant fraction of grand-design spirals in clusters, including those listed in Table~2, are
barred.

Results presented in this paper concern only the cases where the progenitor galaxy is on an exactly prograde orbit
around a cluster. One may wonder whether these results would be applicable for other inclinations between the galaxy
disk's angular momentum and galaxy orbital angular momentum. To clarify this issue to some extent we used additional
simulations already at our disposal performed for orbits O2 and O3 where the initial inclination of the progenitor's
disk was exactly retrograde, $i=180^{\circ}$. In this case no well-defined spiral arms form and the overall effect of
tidal interactions is very mild.

The dependence of the tidal effects on the disk inclination has been recently
addressed by {\L}okas et al. (2015) in the case of dwarf galaxies orbiting a Milky-Way like host. The analysis of
this scaled-down configurations leads to the expectation that our normal-size progenitor galaxy of the present paper
would also form spiral arms for prograde inclinations from $i=0^{\circ}$ to $i=90^{\circ}$. However, the more inclined
cases would generally produce a more complicated 3D spiral structure and a warped disk. On the other hand, for
inclinations close to retrograde, from $i=90^{\circ}$ to $i=180^{\circ}$, no spiral arms or very weak ones would be
formed.

\subsection{Summary}

In this work we have discussed the scenario for the origin of the grand-design spiral arms in galaxies via tidal
interactions with a galaxy cluster. We used $N$-body simulations of a Milky Way-like galaxy evolving inside a
Virgo-like cluster on a few different orbits. The most important findings of this paper may be summarized as follows:

\begin{itemize}
\item Grand-design, two-armed, logarithmic spiral structure forms in galaxies on each orbit around the
Virgo-like cluster.

\item The formation of spiral arms is triggered during the pericenter passages. Later on the arms wind up and dissipate
with time to be triggered again during the next pericenter. This transient and dynamic behavior is reflected in the
measurements of the pitch angle and the arm strength.

\item The strongest arms form on the tightest orbit, however the most extended orbit produces arms that are winding up
most slowly and are therefore most persistent.

\item The pattern speed of the arms decreases with radius and follows the inner Lindblad resonance indicating that
the arms are kinematic density waves.

\item Among the sample of 201 spiral galaxies in the Virgo cluster we find clear 24 grand-design spirals. Nine of
those objects show no signatures of recent interactions with another galaxy, that could trigger their spiral
structure. The morphologies of a few of these 9 galaxies resemble the morphologies of our simulated galaxies.

\end{itemize}

\section*{Acknowledgments}

This work was supported in part by the Polish National Science Centre under grant 2013/10/A/ST9/00023. We thank L.
Widrow for providing procedures to generate $N$-body realizations for initial conditions. We are grateful to an
anonymous referee for useful comments. This research has made use of the NASA/IPAC Extragalactic Database (NED) which
is operated by the Jet Propulsion Laboratory, California Institute of Technology, under contract with the National
Aeronautics and Space Administration. In this work we also used the SIMBAD database, operated at CDS, Strasbourg,
France. We acknowledge as well the usage of the HyperLeda database (http://leda.univ-lyon1.fr) and the use of Galaxy
Zoo database. The galaxy images were provided by the Sloan Digital Sky Survey.


\begin{thebibliography}{}

\bibitem[{Baba et al.}(2013)]{baba} Baba, J., Saitoh, T. R., \& Wada, K. 2013, ApJ, 763, 46
\bibitem[{Bertin et al.}(1989)]{bertin1} Bertin, G., Lin, C. C., Lowe, S. A., \& Thurstans, R. P. 1989, ApJ, 338, 78
\bibitem[{Bertin \& Lin}(1996)]{bertin2} Bertin, G., \& Lin, C. C. 1996, Spiral Structure in Galaxies, A Density Wave
	Theory (Cambridge, MA: MIT Press)
\bibitem[{Bialas et al.}(2015)]{bialas} Bialas, D., Lisker, T., Olczak, C., Spurzem, R., \& Kotulla, R. 2015, A\&A,
	576, 103
\bibitem[{Binney & Tremaine}(1987)]{bt87} Binney, J., \& Tremaine, S. 1987, Galactic Dynamics (Princeton, NJ:
	Princeton Univ. Press)
\bibitem[{Bottinelli al.}(1983)]{incl} Bottinelli, L., Gouguenheim, L., Paturel, G., \& de Vaucouleurs, G. 1983,
	A\&A, 118, 4
\bibitem[{Byrd \& Valtonen}(1990)]{byrd} Byrd, G., \& Valtonen, M. 1990, ApJ, 350, 89
\bibitem[{Cayatte et al.}(1990)]{vla} Cayatte, V., van Gorkom, J. H., Balkowski, C., \& Kotanyi, C. 1990, AJ, 100, 604
\bibitem[{Comerford \& Natarajan}(2007)]{virgo2} Comerford, J. M., \& Natarajan, P. 2007, MNRAS, 379, 190
\bibitem[{de Vaucouleurs et al.}(1991)]{dV1991} de Vaucouleurs, G., de Vaucouleurs, A., Corwin, H., et al. 1991,
	Third Reference Catalogue of Bright Galaxies (New York, NY: Springer)
\bibitem[{Dobbs et al.}(2010)]{dobbs+} Dobbs, C. L., Theis, C., Pringle, J. E., \& Bate, M. R. 2010, MNRAS, 403, 625
\bibitem[{Dobbs}(2011)]{dobbs} Dobbs, C. L. 2011, MSAIS, 18, 109
\bibitem[{Dobbs \& Baba}(2014)]{review} Dobbs, C., \& Baba, J. 2014, PASA, 31, 35
\bibitem[{D'Onghia et al.}(2013)]{elena} D'Onghia, E., Vogelsberger, M., \& Hernquist, L. 2013, ApJ, 766, 34
\bibitem[{Elmegreen et al.}(1991)]{elmegreen0} Elmegreen, D. M., Sundin, M., Sundelius, B., \&  Elmegreen, B. 1991,
	A\&A, 244, 52
\bibitem[{Elmegreen \& Struck}(2013)]{elmegreen1} Elmegreen, B., \& Struck, C. 2013, ApJ, 775, L35
\bibitem[{Elmegreen \& Struck}(2016)]{elmegreen2} Elmegreen, B., \& Struck, C. 2016, ApJ, 830, 115
\bibitem[{Eneev et al.}(1973)]{eneev} Eneev, T. M., Kozlov, N. N., \& Sunyaev, R. A. 1973, A\&A, 22, 41
\bibitem[{Few et al.}(2016)]{few} Few, C. G., Dobbs, C., Pettitt, A., \& Konstandin, L. 2016, MNRAS, 460, 4382
\bibitem[{Fujii et al.}(2011)]{fujii} Fujii, M. S., Baba, J., Saitoh, T. R., et al. 2011, ApJ, 730, 109
\bibitem[{Ghigna et al.}(1998)]{ghigna} Ghigna, S., Moore, B., Governato, F., et al. 1998, MNRAS, 300, 146
\bibitem[{Grand et al.}(2012)]{grand1} Grand, R. J. J., Kawata, D., \& Cropper, M. 2012, MNRAS, 421, 1529
\bibitem[{Grand et al.}(2013)]{grand2} Grand, R. J. J., Kawata, D., \& Cropper, M. 2013, A\&A, 553, A77
\bibitem[{Gunthardt et al.}(2016)]{m51like} Gunthardt, G., Diaz, R. J., \& Aguero, M. P. 2016,
	AJ, accepted, arXiv:1605.02842
\bibitem[{Hart et al.}(2016)]{GZc} Hart, R. E., Bamford, S. P., Willett, K. W., et al. 2016, MNRAS, 461, 3663
\bibitem[{Holmberg}(1941)]{holmberg} Holmberg, E. 1941, ApJ, 94, 385
\bibitem[{Kenney et al.}(2014)]{kenney} Kenney, J. D. P., Geha, M., J\'{a}chym, P., et al. 2014, ApJ, 780, 119
\bibitem[{Koopmann \& Kenney}(2004)]{koopmann} Koopmann, R. A., \& Kenney, J. D. P. 2004, ApJ, 613, 866
\bibitem[{Lin \& Shu}(1964)]{qss} Lin, C. C., \& Shu, F. H. 1964, ApJ, 140, 646
\bibitem[{Lintott et al.}(2011)]{GZa} Lintott, C., Schawinski, K., Bamford, S., et al. 2011, MNRAS, 410, 166
\bibitem[{Lokas et al.}(2015)]{lo15} {\L}okas, E. L., Semczuk, M., Gajda, G., \& D'Onghia, E. 2015, ApJ, 810, 100
\bibitem[{Lokas et al.}(2016)]{lo16} {\L}okas, E. L., Ebrov\'{a}, I., del Pino, A., et al. 2016, ApJ, 826, 227
\bibitem[{Ma}(2002)]{ma02} Ma, J. 2002, A\&A, 388, 389
\bibitem[{Makarov et al.}(2014)]{LEDA} Makarov, D., Prugniel, P., Terekhova, N., et al. 2014, A\&A, 570, 13
\bibitem[{Martinez-Medina et al.}(2014)]{martinez} Martinez-Medina, L. A., Pichardo, B., Moreno, E., \& Peimbert, A.
	2016, MNRAS, 463, 459
\bibitem[{McLaughlin}(1999)]{virgo1} McLaughlin, D. E. 1999, ApJ, 512, L9
\bibitem[{Merritt}(1984)]{merritt} Merritt, D. 1984, ApJ, 276, 26
\bibitem[{Moore et al.}(1996)]{moore96} Moore, B., Katz, N., Lake, G., Dressler, A., \& Oemler, A. 1996, Nature, 379,
	613
\bibitem[{Moore et al.}(1998)]{moore98} Moore, B., Lake, G., \& Katz, N. 1998, ApJ, 495, 139
\bibitem[{Navarro et al.}(1997)]{nfw97} Navarro, J. F., Frenk, C. S., \& White, S. D. M. 1997, ApJ, 490, 493 (NFW)
\bibitem[{Oh et al.}(2008)]{oh8} Oh, S. H., Kim, W.-T., Lee, H. M., \& Kim, J. 2008, ApJ, 683, 94
\bibitem[{Oh et al.}(2015)]{oh15} Oh, S. H., Kim, W.-T., Mok, H. 2015, ApJ, 807, 73
\bibitem[{Pettitt et al.}(2016)]{pettitt} Pettitt, A. R., Tasker, E. J., \& Wadsley, J. W. 2016, MNRAS, 458, 3990
\bibitem[{Saha \& Elmegreen}(2016)]{se16} Saha, K., \& Elmegreen, B. 2016, ApJ, 826, L21
\bibitem[{Salo \& Laurikainen}(2000)]{salo} Salo, H., \& Laurikainen, E. 2000, MNRAS, 319, 377
\bibitem[{Sellwood \& Carlberg}(1984)]{sc84} Sellwood, J. A., \& Carlberg, R. G. 1984, ApJ, 282, 61
\bibitem[{Sellwood \& Athanassoula}(1986)]{sa86} Sellwood, J. A., \& Athanassoula, E. 1986, MNRAS, 221, 195
\bibitem[{Sellwood \& Binney}(2002)]{sb02} Sellwood, J. A., \& Binney, J. J. 2002, MNRAS, 336, 785
\bibitem[{Sellwood}(2011)]{s2011} Sellwood, J. A. 2011, MNRAS, 410, 1637
\bibitem[{Semczuk \& Lokas}(2015)]{sm15} Semczuk, M., \& {\L}okas E. L., 2016, in
	Proceedings of the XXXVII Meeting of the Polish Astronomical Society, ed. A. R\'o\.za\'nska and M. Bejger
	(Warsaw, Poland: Polish Astronomical Society), 252
\bibitem[{Skrutskie et al.}(2006)]{sizes} Skrutskie, M. F., Cutri, R. M., Stiening, R., et al. 2006, AJ, 131, 1163
\bibitem[{Springel et al.}(2001)]{syw01} Springel, V., Yoshida, N., \& White, S. D. M. 2001, New Astronomy, 6, 79
\bibitem[{Springel}(2005)]{spr05} Springel, V. 2005, MNRAS, 364, 1105
\bibitem[{Steinhauser et al.}(2016)]{arepo} Steinhauser, D., Schindler, S., \& Springel, V. 2016, A\&A, 591, 51
\bibitem[{Struck et al.}(2011)]{struck} Struck, C., Dobbs, C. L., \& Hwang, J.-S. 2011, MNRAS, 414, 2498
\bibitem[{Sundelius et al.}(1987)]{sund} Sundelius, B., Thomasson, M., Valtonen, M. J., \& Byrd, G. G. 1987,
	A\&A, 174, 67
\bibitem[{Toomre \& Toomre}(1972)]{tt72} Toomre, A., \& Toomre, J. 1972, ApJ, 178, 623
\bibitem[{Valluri}(1993)]{valluri} Valluri, M. 1993, ApJ, 408, 57
\bibitem[{van den Bergh}(1998)]{vandenbook} van den Bergh, S. 1998, Galaxy Morphology and Classification
	(Cambridge, UK: Cambridge University Press)
\bibitem[{Vera-Ciro et al.}(2014)]{velaciro} Vera-Ciro, C., D'Onghia, E., Navarro, J., \& Abadi, M. 2014, ApJ, 794, 173
\bibitem[{Vollmer et al.}(1990)]{m91} Vollmer, B., Cayatte, V., Boselli, A., Balkowski, C., \& Duschl, W. J. 1999,
	A\&A, 349, 411
\bibitem[{Wenger et al.}(2000)]{simbad} Wenger M., Ochsenbein, F., Egret, D., et al. 2000, A\&AS, 143, 9
\bibitem[{We\.zgowiec et al.}(2007)]{radioKrk} We\.zgowiec, M., Urbanik, M., Vollmer, B., et al. 2007, A\&A, 471, 93
\bibitem[{Widrow \& Dubinski}(2005)]{wd05} Widrow, L. M., \& Dubinski, J. 2005, ApJ, 631, 838
\bibitem[{Widrow et al.}(2008)]{widrow08} Widrow, L. M., Pym, B., \& Dubinski, J. 2008, ApJ, 679, 1239
\bibitem[{Willett et al.}(2013)]{GZb} Willett, K. W., Lintott, C. J., Bamford, S. P., et al. 2013, MNRAS, 435, 2835
\bibitem[{Yun}(1999)]{m81} Yun, M. S. 1999, in Proc. IAU Symp. 186 Galaxy Interactions at Low and High Redshift, ed.
	J. E. Barnes J. E. and D. B. Sanders (Dordrecht, Netherlands: Kluwer), 81
\end{thebibliography}
\end{document}